\documentclass[aps,prl,showpacs,twocolumn,groupedaddress]{revtex4-1}
\usepackage{graphicx}
\usepackage{dcolumn}

\begin{document}


\title{High pressure phases of crystalline tellurium.}


\author{C. Marini,$^{1,2}$ D. Chermisi,$^1$ M. Lavagnini,$^3$ D. Di Castro,$^4$ L. Degiorgi,$^3$ S. Scandolo,$^5$ P. Postorino.$^1$}
\affiliation{$^1$ Dip. Fisica, Universit\'{a} di Roma ``Sapienza", P.le A. Moro 2, I-00185 Roma, Italy.}
\affiliation{$^2$ European Synchrotron Radiation Facility, 6 Rue Jules Horowitz, BP220, 38043 Grenoble Cedex, France.}
\affiliation{$^3$ Laboratorium f\"{u}r Festk\"{o}rperphysik,ETH-Z\"{u}rich, CH-8093 Z\"{u}rich, Switzerland.}
\affiliation{$^4$  CNR-SPIN and Dipartimento di Ingegneria Meccanica, Universit\`{a}  di Roma ``Tor Vergata'', Via del Politecnico 1, I-00133 Roma, Italy}
\affiliation{$^5$ ``Democritos'' CNR-INFM National Simulation Centre, Trieste, Italy}
\affiliation{The Abdus Salam International Centre for Theoretical Physics, Trieste, Italy}


\date{\today}

\begin{abstract}
A study of high pressure solid Te was carried out at room temperature using Raman spectroscopy and Density
Functional Theory (DFT) calculations. The analysis of the P-dependence of the experimental phonon spectrum
reveals the occurrence of phase transitions at 4 GPa and 8 GPa confirming the high-pressure scenario recently
proposed. The effects of the incommensurate lattice modulation on the vibrational properties of Te is discussed.
DFT calculations agree with present and previous experimental data and show the metallization process at 4 GPa
being due to the development of charge-bridges between atoms belonging to adjacent chains. A first-principles
study of the stability of the 4 GPa phase is reported and  discussed also in the light of the insurgence of
lattice modulation.

\end{abstract}

\maketitle

The remarkable advances in high-pressure (HP) methods and techniques have surprisingly brought to light the
occurrence of incommensurately modulated (IM) crystal structures also in elemental systems under pressure
 \cite{Takemura03,Kume05,Nelmes99, Fujihisa07, Degtyareva05,Hejny03_prl,Loa09}. The modulation, although preserving
 the long-range order, removes the lattice translational symmetry and introduces a new structural degree of
 freedom. Moreover in a variety of elemental systems (e.g.  I, Br, P and VI group elements S, Se, Te) the picture
 is even more intriguing since applied pressure induces a metallization transition preceding the onset of the IM phase,
 which suggests a connection with an electronic instability \cite{Loa09}.

Among the chalcogens the long-standing problem of the sequence and the nature of the HP phases of Te has been
the subject of intense theoretical \cite{Hsueh99,LDA+GCC} and experimental \cite{Hejny03_prl,Bardeen49,Loa09}
investigations. Despite the great effort, a comprehensive picture is still far from being reached at least over
the intermediate pressure range. It is indeed well known that, under ambient conditions and on increasing the
pressure up to 4 GPa, Te is a semiconductor and its crystal structure is trigonal with spiral atom chains along
the c axis (Te-I). The three atoms in the unit cell are linked via covalent-like bonds to the nearest neighbors
along the chain and, via Van der Waals interactions, to the second neighbors lying along adjacent chains
\cite{Aoki80}. The very HP regime (i.e. above 27 GPa or 29 GPa according with Ref.\cite{Parthasarathy88} or
Ref.\cite{Hejny03_prl} respectively) is established, as well: Te is a metal with a highly symmetric bcc
structure (Te-IV).

Apart from the metallization transition that, since long, it is known to occur at 4 GPa \cite{Bardeen49}
together with a structural change, there is an intense debate on the HP metallic structural phases between the
Te-I and the Te-IV. i.e. over the 4-27(29) GPa pressure range. In particular the succession of phases over this
intermediate pressure range accepted up to few years ago was: a puckered layer monoclinic structure for P $>$ 4
GPa which transforms into an orthorhombic one at 6.8 GPa and finally a $\beta$-Po-type above 11 GPa and below
the Te-IV pressure threshold \cite{Aoki80,Jamieson65,Parthasarathy88}. Recent X-ray diffraction studies
\cite{Hejny03_prl,Hejny04_prb} have drawn a rather different HP phase transition scenario, claiming a structural
transition to triclinic (Te-II) at 4.0 GPa, followed by the onset of an IM monoclinic phase (Te-III) at 4.5 GPa
which is finally transformed into the bcc Te-IV phase above 29 GPa. It is worth to notice that since Te-I is
observed up to 4.5 GPa and Te-II up to 8 GPa, we have large regions of coexisting phases and the above reported
phase boundaries are intended as the pressure threshold at which the higher pressure phase first appears on
compressing the lattice.

In the present paper a combination of experiments and \textit{ab initio} calculations is used to study
structural, dynamic and electronic properties of HP solid Te in order to solve the controversy about the
sequence of crystallographic structures, the real occurrence of IM phases, and to shed light on the microscopic
metallization mechanism. A HP Raman study of Te (0 $-$ 15 GPa) is reported together with the results of density
functional theory (DFT) calculations which well reproduce the experiment and allow to gain a further insight of
the microscopic modifications induced by pressure.

\begin{figure}[h!]
\includegraphics[height=12.5cm]{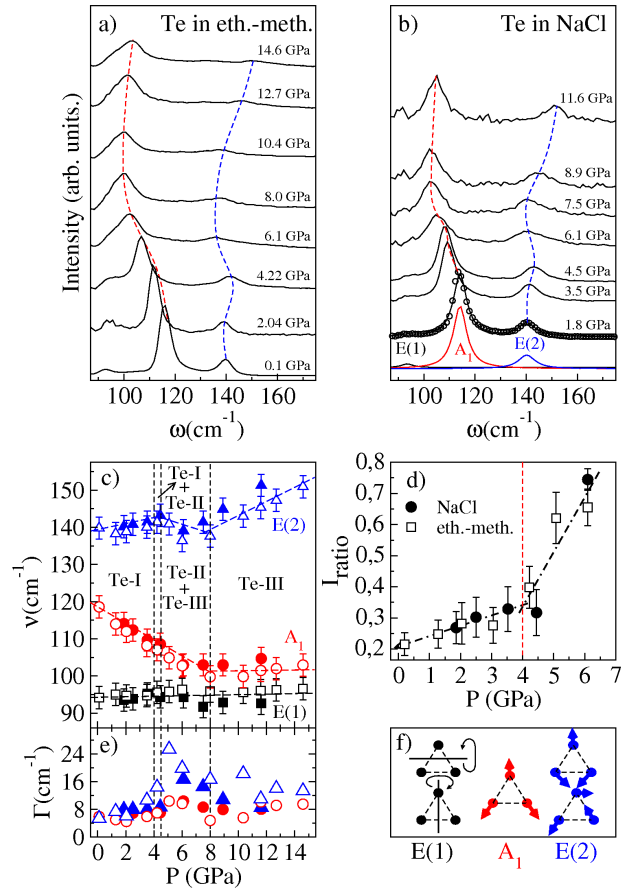}
\caption{\label{all} (color online) Te Raman spectra at RT and selected P, in ethanol-methanol (a), and NaCl
(b). Best fit curve and phonon contributions for the P=1.8 GPa spectrum are also shown in (b). P-dependence of
phonon frequencies (c) and linewidths (e). Vertical lines in (c) and (e) mark the phase transitions pressures
according to ref. \cite{Hejny04_prb}. The intensity $I_{ratio}= I[E(2)]/(I[E(1)] + I[A_1])$ (see text) vs.
pressure in (d). Open (close) symbols in (c), (d) and (e) refer  to measurements using  ethanol-methanol mixture
(NaCl). Color code and atomic displacements for the Raman active modes in Te-I (f). }
\end{figure}

A high purity Te sample (99.99\% by Aldrich) was used without further treatment. A diamond anvil cell coupled to
a microRaman spectrometer (Horiba Jobin Yvon) was employed. The ruby fluorescence technique was applied for
pressure calibration and two pressure-transmitting media (NaCl and methanol/ethanol 1:4 mixture) were used to
verify their possible effects on the high pressure measurements. Moreover, exploiting the high spatial
resolution of our experimental setup, Raman spectra were collected at each pressure from different points over
the surface of the samples (about 50$\times$50 $\mu$m$^2$) to test the presence of pressure gradients. Further
experimental details can be found in Ref. \cite{NOI}, here we just notice that great care was taken in the
optical alignment \cite{Degiorgi} to collect reliable spectra above 80 cm$^{-1}$ and that we did not find any
evidence of either reactions between the media and the sample or detectable non-hydrostaticity  effects. No
laser-induced sample heating has been observed as well\cite{Paolone2005635}. Normalized and background
subtracted (see below) Raman spectra of Te collected using NaCl and methanol/ethanol are shown in Fig. \ref{all}
at selected P values. According to the standard group theory (GT), the irreducible representation of the optical
modes of Te-I is A$_1$ + A$_2$ + 2 E. The Raman active modes are: the A$_1$ mode (breathing in the \textit{ab}
plane) and the two doubly degenerate E modes (E(1) \textit{a-} and \textit{b-}axis rotation , E(2) asymmetric
stretching mainly along \textit{c-}axis). These latter are also Infrared active since the crystal lacks a center
of inversion. Measured spectra were fitted by using a standard model curve \cite{fit} given by the sum of three
phonon contributions (each described by a damped harmonic oscillator), and an electronic background. A linear
baseline was also included in the fitting curve. In Fig. \ref{all}b the best-fit curve and the phonon assignment
are also shown for the lowest pressure spectrum. The spectra shown in Fig. \ref{all}a and Fig. \ref{all}b are
background subtracted and normalized to the phonon integrated intensities.

The Raman spectrum shows a remarkable pressure dependence albeit the three-peaks structure is basically
preserved over the whole pressure range. We notice that the persistence of a Raman signal well above 11 GPa
disagrees with the old phase transition scheme \cite{Aoki80,Jamieson65,Parthasarathy88} since no Raman-active
are expected for the $\beta$-Po-type phase. On the other hand, an unmodulated monoclinic phase (i.e. an
unmodulated Te-III structure) has no Raman-active mode as well. The observation of a Raman signal is thus only
coherent with the new transition scheme where the Te-III structure appears to be IM \cite{Hejny03_prl}. In this
case, optical scattering processes are allowed not only from phonons at $\Gamma$ but, in the sinusoidal
approximation, also at $k= \pm q_i$  (with $q_i$ the modulation wave vector). Using the superspace GT (equations
and formalism are from Ref. \cite{Janssen79_jpc}), the calculated character of the representation is
$\chi(E,2_y,m_y,I)=(6,2,0,0)$ where E, 2$_y$, m$_y$ denote the point group operations of the three-dimensional
space-group elements. This representation is reducible into $\Gamma$ = 2 A$_g$ + 1 B$_g$ + 2 A$_u$ + 1 B$_u$,
where three Raman active modes (two A$_g$ modes and the B$_g$ mode) are expected. This first experimental
evidence thus provides not only a clear support to the recently proposed transition scheme but can be seen as a
marker of the onset of an IM phase.

Looking at Fig. \ref{all}c, three different pressure regions can be identified by the abrupt variations of the
pressure dependence of the phonon peak frequencies. The borders among the regions actually correspond to the
pressure thresholds of the new transition scheme \cite{Hejny03_prl,Hejny04_prb}. Over the 0-4 GPa pressure
range, the phonon frequency of  the A$_1$ mode shows a linear softening ($\sim$ 10 cm$^{-1}$), whereas the
frequencies of the two E modes weakly increase.  Since the A$_1$ mode is a chain breathing and, over the same
pressure range, a contraction of the \textit{b} lattice parameter and an almost constant \textit{c}
\cite{Aoki80} were observed, the above findings suggest a progressive weakening of intrachain bonds in favor of
interchain atomic interactions. The slope discontinuity of the pressure dependence of the E(2) mode whose
frequency starts to decrease at 4 GPa, marks the emergence of a new crystal phase albeit, within the 4-7 GPa,
the frequency of the A$_1$ phonon is still decreasing and that of the the E(1) mode keeps almost constant. The
narrow pressure range in which Te-I and Te-II coexist (4-4.5 GPa) makes it impossible to determine whether the
discontinuity can be ascribed to the Te-I/Te-II transition or to the onset of the Te-III phase. According to
 Ref.~\cite{Hejny03_prl,Hejny04_prb} that reports on the Te-II/Te-III coexistence within the 4.5-8  GPa pressure
range, the observed softening of the E(2) mode can be ascribed to the structural instability of the Te-II phase
which is progressively converted into the rather similar Te-III phase \cite{Hejny03_prl,Hejny04_prb}. We notice
that larger values of the phonon linewidths shown in Fig. \ref{all}e can be found within the coexistence region
(4.5-8  GPa) as a consequence of the lattice disorder. On further increasing the pressure above $\sim$ 8 GPa all
the phonon modes show a rather regular frequency hardening, indicating that a single, stable phase (Te-III) has
finally established.

The presence of modulation implies a one-to-one correspondence between eigenvalues
 of the dynamical matrix at generic  $k$
and those at $k\pm q_i$.\cite{Janssen79_jpc}
More precisely the phonons at $\Gamma$
involved in the extra scattering processes arising from the modulation,
have the same energies as the phonons  at the point $q_i$ involved in
 the usual inelastic scattering processes.\cite{Janssen79_jpc}
This allows an unusual comparison between the results of a Raman experiment and those of an inelastic x-ray
(neutron) scattering experiment. Indeed, the frequency of the Raman mode at higher energy, observed at 8.9 GPa
(Te-III phase, see Fig. \ref{all}c), is the same within the experimental uncertainties as the frequency  of the
longitudinal mode at $k=q_i$ observed by the inelastic x-ray scattering experiment reported in
Ref.~\cite{Loa09}. This finding thus provide further evidence of the the onset of an IM phase at HP.

\begin{figure}[h]
\includegraphics[width=8.0cm]{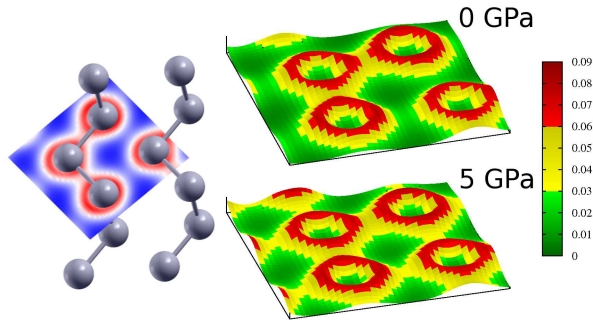}
\caption{
 (color online) Valence electron density at 0 and 5 GPa in the Te-I
phase over  the plane containing 3 atoms along the chain and the closest atom in the adjacent chain.}
\label{rho}
\end{figure}

The weakening of intrachain bonds can be ascribed to the onset of a pressure induced charge transfer process, as
also suggested by the remarkable reduction of the intensity (peak integrated area) of the A$_1$ mode, I[A$_1$]
~\cite{iodio}. This effect is shown in Fig. \ref{all}d where the pressure dependence of the intensity ratio
$I_{ratio} = I[E(2)]/(I[A_1] + I[E(1)])$ is shown.  Neglecting the small contribution to $I_{ratio}$ due to the
small intensity (area) of the E(1) mode, the growth of $I_{ratio}$ indicates a strong decrease in the intensity
of the A$_1$ peak before the metallization pressure. Moreover two different regimes, below and above the
metallization pressure (4 GPa), can be clearly identified.

To gain further insight into the microscopic charge-transfer process, pseudopotential DFT calculation using the
Quantum-Espresso code \cite{espresso} have been carried out. A full structural relaxation under the influence of
Hellmann-Feynman forces and stresses was carried out for Te-I at 0, 2.5 and 5 GPa using the variable cell shape
method\cite{Hellmann-Feynman}. At each pressure the electronic density of states and bands as well as the
frequencies and the intensities of normal vibrational modes were obtained by the perturbation approach of DFT
\cite{perturbation}. We found the closure of the energy gap at 5 GPa and an overall good agreement between the
calculated data and the HP experimental structural data \cite{Aoki80}. The intrachain bond length was slightly
overestimated because of the typical error of the generalized gradient approximations in the determination of
the cohesive force in strongly anisotropic systems.
This reflects in an underestimate of the calculated phonon frequencies (by less than 10\%) which, however,
together with the intensity ratio $I_{ratio}$ defined above, shows the same pressure dependence as the
experimental data. Moreover, the calculations  support the evidence for a progressive reduction of the angle
between the E(2) mode eigenvectors and the \textit{c-}-axis on applying pressure (from 12$^{\circ}$ at P=0 to
2.5$^{\circ}$ at P=5). The good agreement of the results of DFT calculations with the present and the previous
experimental data makes us confident also about the other possible outputs of the code. In particular our
calculations provided also the maps of the valence electron (5S$^2$5P$^4$) densities shown in Fig. \ref{rho} at
ambient pressure and at 5 GPa . The onset of a bridge of charge density linking the atoms inside the chain and
the first neighbor atom in the adjacent chain is pretty clear at P=5 GPa. The pressure induced metallization
transition is therefore the result of the simultaneous lattice symmetrization and charge-transfer processes from
intrachain to interchain regions which allows the mobility of the valence electron.

\begin{figure}[h!]
\includegraphics[height=5cm]{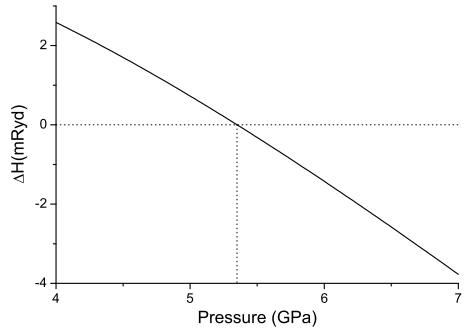}
\caption{The difference $\Delta H= H_{Te\textrm{-}II}-H_{Te\textrm{-}I}$ between the enthalpy values of the
Te-II and Te-I  cells relaxed  at different pressures using the variable cell shape method. Dotted lines mark
the crossover.
}
\label{entalpy}
\end{figure}

We also exploited the DFT approach to investigate the structural Te-I/Te-II transition although it has proved to
be a rather complex task. In Te-I most of the crystallographic constants are fixed by the hexagonal geometry,
nevertheless the DFT underestimates the anisotropy of the system leading to a slight overestimation of the ratio
of chain radius to interchain distance. In the Te-II case, where all the lattice parameters and the atom
positions are
 left free to vary during the cell relaxation, this trend is strongly emphasized, so that, with relaxing the structure,
 the experimental zig-zag puckering is entirely lost. For this reason, the  dynamic properties for the Te-II relaxed
cells in the pressure range between 3.5 and 7 GPa differ qualitatively from the experimental ones, even if the
calculated enthalpy of the Te-I phase becomes higher than that of  Te-II phase above 5.5 GPa, in agreement with
the occurrence of the phase transition (see Fig. \ref{entalpy}).
Thus, in the present case, the dynamical properties were calculated
optimizing  the atomic positions while keeping fixed the lattice parameters
to the experimental values obtained for the Te-II phase at 4.5 GPa \cite{Hejny04}.
For sake of comparison the Raman frequencies have been recalculated
by optimizing only the atomic positions keeping constant the lattice
parameters to the experimental values observed  for the Te-I phase at 4.0 GPa.
 The results for the two phases are reported in Tab. \ref{DFT_Raman}.
\begin{table}[!ht]
\caption{\label{DFT_Raman} Calculated Raman phonon frequencies for Te-I and Te-II optimizing the atom positions
while keeping constant the lattice parameters to the experimental values.} \centering

\begin{tabular}{|c|c|c|c|}
\hline
\multicolumn{2}{|c|}{Te-I at 4 GPa}&\multicolumn{2}{|c|}{Te-II at 4.5 GPa}\\
\hline
Mode        & $\nu$     (cm$^{-1}$) & Mode        & $\nu$     (cm$^{-1}$)\\
E(1)        &       78              &   A$_g$(1)  &     38               \\
A$_1$       &       92              &   A$_g$(2)  &     94               \\
E(2)        &      122              &   A$_g$(3)  &    125              \\
\hline
\end{tabular}
\end{table}

We notice that only the low-frequency phonon shows a remarkable discontinuity at the transition whereas the
differences between the values calculated in the two phases for the high-frequency modes are within the typical
DFT calculations accuracy. In agreement with the experimental data our calculations thus predict a continuous
evolution of the frequencies across the Te-I Te-II transition for the high-energy modes. As to the estimate of
the frequency of the low-energy mode further discussion is required. This vibrational mode is the mode mostly
affected by the IM since the the direction of the incommensurate  modulation is rather close to the direction
along with the atom displacements occur The underestimate of the mode frequency seems to be related to an
excessive sensitivity of the DFT  theory to predict the incommensurate phase, or in other words it seems to be
related to an  overestimate of the width of the saddle point of the energy landscape in the direction of the
modulation. This is confirmed by the fact that there is a neighborhood of points around q$_{inc}$ for which the
transverse acoustic phonon has negative energies \cite{Hejny03_prl}.

In summary the Raman experiment supports the recently proposed HP phase transition scenario \cite{Hejny03_prl}
and the appearance of an incommensurate modulation is clearly demonstrated at intermediate pressure. Our DFT
calculations well reproduce the present and the previous HP experimental data allowing for an accurate study of
the microscopic pressure induced modifications. In agreement with the indications of the Raman experiment, DFT
calculations show that the metallization transition is driven by an intra- to inter-chain charge transfer which
develops charge density bridge among adjacent chains. Finally the combined use of spectroscopic and \textit{ab
initio} methods has proved to be a useful approach to the complexity of systems with incommensurately modulated
structure.

\end{document}